\documentclass [prb,twocolumn,notitlepage,superscriptaddress,floatfix] {revtex4-1}
\usepackage{amsmath}
\usepackage{graphicx}
\usepackage{lmodern}
\usepackage{amsmath}

\usepackage{color}
\usepackage{amssymb}
\newcommand{\beq} {\begin{equation}}
\newcommand{\eeq} {\end{equation}}
\newcommand{\bea} {\begin{eqnarray}}
\newcommand{\eea} {\end{eqnarray}}
\newcommand{\be} {\begin{equation}}
\newcommand{\ee} {\end{equation}}
\renewcommand{\(}{\left(}
\renewcommand{\)}{\right)}
\renewcommand{\[}{\left[}
\renewcommand{\]}{\right]}

\DeclareMathOperator{\Tr}{Tr}

\newcommand{\so} {{\mathrm{SO(4)}}}

\newcommand{\z} {{\mathbb Z_2}}

\begin{document}

\title {Interplay between uni-directional and bi-directional charge-density-wave orders in underdoped cuprates }
\author{Yuxuan Wang}
\affiliation{Department of Physics and Institute of Condensed Matter Theory, University of Illinois at Urbana-Champaign, Urbana, Illinois 61801, USA }
\author{Andrey V. Chubukov}
\affiliation{School of Physics and Astronomy and William I. Fine Theoretical Physics Institute,
University of Minnesota, Minneapolis, MN 55455, USA}
\date{\today}

\begin{abstract}
We analyze the interplay between charge-density-wave (CDW) orders
with axial momenta $(Q,0)$ and $(0,Q)$ ($\Delta_x$ and $\Delta_y$  respectively),  detected in the underdoped cuprates.
 The CDW order in real space can be
 uni-directional (either $\Delta_x$ or $\Delta_y$ is non-zero) or bi-directional (both $\Delta_x$ and $\Delta_y$ are non-zero).  To understand which of the two
  orders develop, we adopt
   the magnetic scenario, in which the CDW order
    appears due to spin-fluctuation exchange,
     and derive the Ginzburg-Landau action to the sixth order in $\Delta_x$ and $\Delta_y$.
 We argue that, at the mean-field level,
  the CDW order is bi-directional at the onset, with equal amplitudes of $\Delta_x$ and $\Delta_y$,
   but changes to uni-directional inside the CDW phase.  This implies that, at a given temperature,
 CDW order is uni-directional at smaller dopings, but
   becomes bi-directional at larger dopings.
    This is consistent with recent x-ray
     data on YBCO,
     which detected tendency towards bi-directional order at larger dopings.
  We discuss the role of discrete symmetry breaking at a higher  temperature for the interplay between bi-directional and uni-directional CDW orders
   and also discuss
    the role of pair-density-wave (PDW) order, which may appear along with
   CDW.  We argue that
  PDW with the same momentum as CDW
  changes the structure of the bi-directional charge order by
   completely
    replacing
   either $\Delta_x$ or $\Delta_y$ CDW components by PDW.
   However, if an
   ``Amperean"
   PDW order,
        which pairs fermions with approximately the same momenta,
    is also present,
     both $\Delta_x$ and $\Delta_y$ remain non-zero
     in the bi-directional phase,
      albeit with non-equal amplitudes.  This is again consistent with x-ray experiments, which at larger doping found non-equal $\Delta_x$ and $\Delta_y$ in every domain.
\end{abstract}
\maketitle

\section{Introduction}

 Understanding of the charge-density-wave (CDW) order in high-$T_c$ cuprates is an essential step  towards the
 understanding of the phase diagram of these materials.
  An incommensurate CDW order has been observed in La-based cuprates a while ago~\cite{tranquada,tranquada1}, and recently was found to be ubiquitous in the
  cuprates~\cite{ybco,ybco_2,ybco_1,x-ray, x-ray_1,x-ray_2,x-ray_last,keimer,davis_1,mark_last,david}. The CDW order is incommensurate,
  with momentum ${\bf Q}$  along $X$ and/or $Y$  directions,
  where $Q\sim (0.2-0.3)\times 2\pi$.
The charge order observed in zero magnetic field is static but short-ranged (probably pinned by impurities~\cite{mark_last}).
 In a finite  field a true long-range CDW order has been detected~\cite{long_range}.  An incommensurate  charge order parameter generally has
  both on-site  and bond components (a true CDW and a bond order, respectively~\cite{davis_1,x-ray_2}.
  To simplify the presentation, we use the term CDW below for both on-site and bond orders.

The presence of the two axial momenta
$Q_x=(Q,0)$ and $Q_y =(0,Q)$, and, hence, two distinct U(1) components $\Delta_{Q_x} = \Delta_x$ and $\Delta_{Q_y} = \Delta_y$, naturally raises the
  question
  whether both are present simultaneously in the CDW state, or only one
   component orders~\cite{charge,debanjan,akash,stripe_eduardo}. If both $\Delta_x$ and $\Delta_y$ are
present, the
 CDW order is called bi-directional.
 If the amplitudes of $\Delta_x$ and $\Delta_y$ are equal, bi-directional CDW order does not break $C_4$ lattice rotational symmetry.  If
  only $\Delta_x$ or only $\Delta_y$ develops, the order is uni-directional, and in the ordered phase the system breaks not only $U(1)$ translational symmetry, but also
  $C_4$ symmetry down to $C_2$,  by spontaneously choosing $\Delta_x$ or $\Delta_y$.
   At the mean-field level, $C_4$ and $U(1)$ symmetries get broken at the same $T$.
     Beyond mean-field, the $C_4 \to C_2$
   symmetry breaking occurs at a higher $T$
than the breaking of a continuous U(1) symmetry, and this gives rise to a nematic
  state
  at intermediate $T$'s,
   in which the rotational
    $C_4$ symmetry is broken down to $C_2$,
     but the translational U(1) is preserved~\cite{charge,rafael_nem}.  The CDW order,
      either uni-directional or bi-directional,
      may also break $Z_2$ time-reversal symmetry, if the phases of the CDW orders with the same ${\bf Q}$ but opposite center of mass momenta are not identical~\cite{charge,tsvelik,rahul}. We discuss one such state below.

Recent X-ray and STM experiments on underdoped cuprates \cite{davis_1,x-ray_last,keimer} point towards a uni-directional CDW,
 also known as the ``stripe order"~\cite{tranquada,fluct_stripe}.  However,  X-ray data on YBCO  at larger dopings were interpreted~\cite{keimer,andrea}  as evidence that at higher hole
  concentration  the order switches from uni-directional to bi-directional.
Specifically, at at lower dopings resonant x-ray scattering data show
only one peak at momenta $Q_x$ and $Q_y$ in every domain, while at
 higher dopings two peaks at momenta $Q_x$ and $Q_y$ have been detected in every domain, with unequal intensity. The difference between the intensities was interpreted to be due to intrinsic orthorhombicity.  The bi-directional CDW order was also assumed in the interpretation of quantum oscillations in a magnetic field~\cite{suchitra}.

In this paper we analyze the interplay between CDW order parameters with momenta $Q_x$ and $Q_y$ within the spin-fluctuation scenario~\cite{acs,ms,efetov,ms,charge}.
 In this scenario, axial CDW order with predominantly $d$-wave form factor emerges in a paramagnetic state due to effective attractive interaction mediated
 by soft spin fluctuations peaked at or near $(\pi,\pi)$, much like spin-mediated $d$-wave superconductivity.
   Within a given ``hot region" in the Brillouin zone, the spin-fluctuation exchange gives rise to
   CDW order with a small momentum, much like as the one due to small-$Q$
      phonon exchange~\cite{phonon},
    Magnetically mediated CDW order
     also naturally gives rise to a sign change between
         CDW orders in different hot regions separated by $(\pi,\pi)$.
         This is
          consistent with the observed $d$-wave
          form-factor of the CDW order parameter~\cite{davis_1}.

          In this paper we consider the clean system, in which CDW order emerges as a true long-range order at $T=0$ and as an algebraic order with power-law decays of correlations at a finite $T < T_{BKT}$. In the real materials, a CDW order is likely pinned by impurities~\cite{nie} and is short-range, albeit static.

 We derive the Ginzburg-Landau Free energy to sixth order in CDW order parameters $\Delta_x$ and $\Delta_y$. These two order parameters couple to fermions in hot regions on the FS, and the coefficients in the Free energy are given by loop diagrams made out of hot fermions.
 We compare
 mean-field
  Free energies of uni-directional and bi-directional CDW orders
 and argue that,
 at its onset, the CDW order is bi-directional.
  However, the order
   changes to uni-directional inside the CDW-ordered phase, once
  the magnitude of CDW order parameter
  exceeds some critical
   value.
    Since the onset temperature $T_{\rm CDW}$ is a decreasing function of doping $x$, the
   CDW order,
   viewed as a function of doping at a given temperature,
   is uni-directional at smaller dopings and bi-directional
    at higher dopings.  This result is consistent with recent x-ray experiments on YBCO~\cite{keimer,andrea}, which, as we said,  were interpreted  as evidence that at higher hole
  concentration  the order switches from uni-directional to bi-directional.

We argue that the mean-field Free energies of uni-directional and bi-directional CDW immediately below $T_{\rm CDW}$
differ substantially at small $T_{\rm CDW}$ but become rather close to each other at higher
$T_{\rm CDW})$. In this last case, uni-directional CDW may win over bi-directional CDW already from the onset, once we go beyond mean-field and  include nematic fluctuations, which favor uni-directional CDW.  We show the most likely phase diagram in Fig. 1.

For completeness we also analyze how the doping evolution of the CDW order is affected by potential presence of  the   pair-density-wave (PDW) order.
 This order has
 been proposed in several theory papers~\cite{patrick,agterberg,kivelson,pdw_frad,pdw,pdw2,pepin} and was recently reported to be observed in the tunneling experiments on the cuprates~\cite{davis_last}.

 In the spin-fluctuation scenario, a PDW order with the same $Q_x$ and/or $Q_y$ appears to be almost degenerate with CDW order~\cite{pdw,pdw2,pepin}
 due to approximate particle-hole SU(2) symmetry~\cite{ms,efetov}. The presence of such PDW does not affect qualitatively the uni-directional phase as CDW/PDW order still develops with the (relative) momentum $Q_x$ or $Q_y$, but
 it does affect the structure of CDW in the
 bi-directional phase. Namely,
   CDW develops along one direction, say with the relative momentum $Q_x$,
    and  PDW order develops along the orthogonal direction
    with the relative momentum
    $Q_y$ (Refs. \ \onlinecite{pdw,pdw2}).
     Such a structure would still show up as uni-directional in the experiments which probe  only CDW component, in disagreement with the X-ray data~\cite{keimer,andrea}.   We argue, however, that the consistency with X-data can be restored if the system also develops PDW order involving fermions from the same hot region, as Refs.~\onlinecite{patrick,agterberg} suggested.  Such an order mixes particles and holes within a given hot region and, as a result,
       CDW component appears with $Q_x$ and with $Q_y$, albeit with non-equal magnitudes.

  The structure of the paper is the following. In Sec. II we discuss the model. In Sec. III we assume that only CDW order develops,
   and analyze the structure of CDW order at the onset and inside the CDW-ordered phase.  In Sec. IV we consider potential co-existence of CDW and PDW orders.
   Sec. V presents the summary and the conclusions.

   It is instructive to place our work in the context of other studies of the interplay between uni-directional and bi-directional CDW/PDW order in doped cuprates.
   The structure of CDW order without PDW  has been analyzed before~\cite{charge,tsvelik,debanjan}, but only at its onset and at the lowest $T$.  In this work we extend the analysis of CDW order at the onset to larger $T$, and also analyze the structure of CDW order inside the ordered phase.
      The co-existence of PDW and CDW orders
    with $Q_x$ and $Q_y$ immediately below the CDW/PDW instability has been analyzed in Refs.\ \onlinecite{pdw,pdw2,pepin}, again at small $T$. The
    PDW order with total momenta approximately equal to twice hot spot value was considered  in Ref.~\onlinecite{patrick} without reference to hot spot scenario and in Ref.\ \onlinecite{agterberg} within the hot spot model. (In the latter case the total momentum of a pair is actually along one of Brilliouin zone diagonals, i.e. it is $Q_{\rm diag} = (Q, \pm Q)$, because hot spots are located at the intersection with magnetic Brillouin zone boundary.)
     We analyze the interplay between $Q_x/Q_y$ and $Q_{\rm diag}$ orders when both are present.

\begin{figure}
\includegraphics[width=\columnwidth]{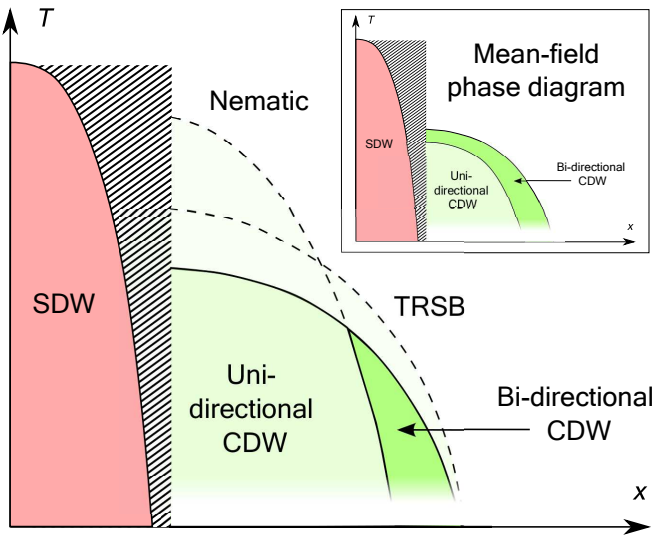}
\caption{The schematic phase diagram for the CDW order.
Inset: the CDW phase diagram obtained within the mean-field theory. 
The CDW order is bi-directional at the onset but becomes uni-directional inside the ordered phase.
 The main figure: The full phase diagram beyond mean-field theory. The uni-directional order still appears inside the CDW-ordered phase at low temperatures, but 
  becomes the only order at higher temperatures due to feedback effect from the nematic order  which sets up prior to CDW order.
The two dashed curves correspond to the onset of the nematic order and of time-reversal-symmetry-breaking. The latter occurs independent on whether CDW order is uni-directional and bi-directional.  In the shaded region, Mott physics develops and the range of charge ordering shrinks.   This phase diagram implies that bi-directional order is completely eliminated once the onset temperature of the $Z_2$ nematic order exceeds that of $U(1)$ CDW order. Another possibility (not shown) is that bi-directional order survives below the nematic transition line, but the magnitudes of CDW orders $\Delta_x$ and $\Delta_y$ become non-equivalent, in line with the breaking of
  $C_4$ symmetry.}
\end{figure}

\section{The model}

We follow earlier works and consider two-dimensional metallic system with
 the Fermi surface
 shown in Fig.\ 2.
We define CDW order parameters $\Delta_x$ and $\Delta_y$
 as
$\Delta_{{\bf k}_j}^{i}=\sum_{k \approx k_j} \langle c^\dagger_{{{\bf k+Q}_i}/2} c_{{{\bf k-Q}_i}/2} \rangle$, where $i = x,y$ and the summation over center of mass momentum $k$ is restricted to the vicinity of one of
 eight
 ${\bf k}_j$ points,
 for which ${\bf k}_j \pm {\bf Q}_i/2$ are both at the Fermi surface
 ( see Fig.\ 2, these points are often called the hot spots). The
  momenta ${\bf k}_j$  are not high symmetry points
   in the Brillouin zone~\cite{charge,agterberg},
    hence
     $\Delta_{k_j}^i$ and $\Delta_{-k_j}^i$ are
     generally not identical, despite that they have the same ${\bf Q}_i$.
     If spin fluctuations are peaked at $(\pi,\pi)$,
  then
    ${\bf k}_j$ is along $X$ direction for ${\bf Q} = {\bf Q}_y$ [${\bf k}_j = \pm k_x = (\pm (\pi- Q),0)$] and
    ${\bf k}_j$ is along $Y$ direction for ${\bf Q} = {\bf Q}_x$ [${\bf k}_j = \pm k_y = (0,\pm (\pi- Q))$].

 We
 label hot regions in Fig.\ 2 as $\pm 1, \pm2, \pm3, \pm4$ and define the
 Fermi velocity at hot spot 1 as $(v_x,v_y)$, the one at hot spot 2 as $(v_x,-v_y)$, etc.
 The magnitude of the velocity $v =\sqrt{v^2_x + v^2_y}$ is the same for all hot spots.
 The fit of ARPES data for Bi2212 by tight-binding dispersion yielded~\cite{norman} a large ratio
  of velocities $v_y$ and $v_x$: $v_x/v_y = 13.6 $. We use this as an input and
  set  $v_y\gg v_x$ in our calculations.
  The fermionic dispersion $\epsilon_{i,\tilde k}$ near a given hot spot $i$ is linear in momentum deviation $\tilde k$ from the hot spot, e.g., $\epsilon_{1,\tilde k}=v_x \tilde k_x +v_y\tilde k_y$, $\epsilon_{2,\tilde k}=v_x \tilde k_x -v_y\tilde k_y$, etc.
  We assume that  the linear dispersion holds up to energy scale $\Lambda$ which we set as the upper cutoff in our low-energy theory.

\begin{figure}
\includegraphics[width=0.8\columnwidth]{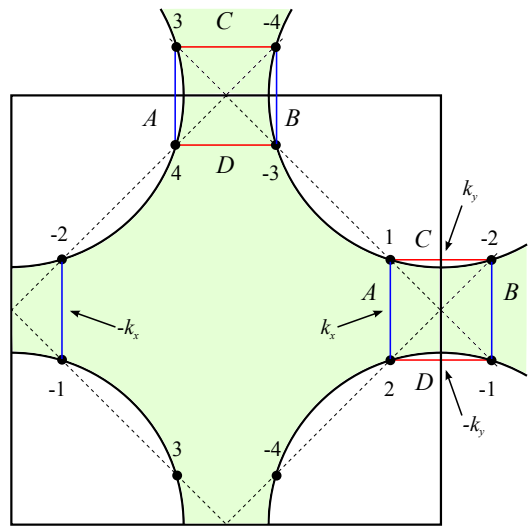}
\caption{The Brillouin zone (BZ) of a typical cuprate system and the hot spots $\pm (1,2,3,4)$, which are defined as points on the Fermi surface separated by the antiferromagnetic momentum ${\bf K}=(\pi,\pi)$. The Fermi velocity at a hot spot, say 1, is given by ${\bf v}=(v_x,v_y)$. The CDW order parameters, labeled by $A,B,C,D$, couples with hot fermion pairs in the axial direction. CDW order parameters with the same momentum, for example $\Delta_A$ and $\Delta_B$, are not equivalent since they have opposite center-of-mass momentum $\pm k_0$.}
\end{figure}

 \section{Uni-directional vs bi-directional CDW order}

 We first assume that only CDW order develops  and derive the Free energy for four CDW order parameters $\Delta_{A,B}=\Delta^{Q_y}_{\pm k_x}$ and $\Delta_{C,D}=\Delta^{Q_x}_{\pm k_y}$ to order $\Delta^6$.

The CDW order parameters couple to hot fermions via
 \begin{align}{\mathcal H}'=&\Delta_A[c_2^\dagger(\tilde k) c_1(\tilde k)-\mu c_4^\dagger(\tilde k) c_3(\tilde k)]\nonumber\\
 &+\Delta_B[c_{-1}^\dagger(\tilde k) c_{-2}(\tilde k)- \mu c_{-3}^\dagger(\tilde k) c_{-4}(\tilde k)]\nonumber\\
 &+\Delta_C[\mu c_{1}^\dagger(\tilde k) c_{-2}(\tilde k)- c_3^\dagger(\tilde k) c_{-4}(\tilde k)]\nonumber\\
 &+\Delta_D[\mu c_2^\dagger(\tilde k) c_{-1}(\tilde k)- c_4^\dagger(\tilde k) c_{-3}(\tilde k)]+h.c.,
 \label{zz}
 \end{align}
 where the minus sign within each bracket accounts for
  the sign change of CDW order under ${\bf k}_j \to {\bf k}_j + (\pi,\pi)$
   and $\mu >1$ describes the difference in magnitudes between CDW orders between, e.g., hot spots $1-2$ and $3-4$ in Fig. 2: $\Delta_{1-2} = \Delta_A$,
    $\Delta_{3-4} = - \mu \Delta_A$ (for details see Refs.\ \onlinecite{charge, debanjan}).  When $\mu =1$, CDW order has a pure $d-$wave form, when $\mu$ differs from one, it has both $d-$wave and $s-$wave components ($\Delta_A (1+\mu)$ and $\Delta_A (\mu-1)$, respectively). The $d-$wave component is always larger.
      The model calculations of Refs.  \ \onlinecite{charge, debanjan} yield $\mu = \sqrt{\log{(v_y/v_x)}}$.

 \subsection{Selection of CDW order near its onset}

 The Free energy
 in terms of $\Delta$ is obtained by integrating out fermions in the partition function for ${\cal H}$ given by the sum of free-fermion Hamiltonian and  ${\mathcal H}'$ from
 (\ref{zz}) and re-expressing the result as $\int d \Delta e^{-F_{CDW}/T}$. Expanding $F_{CDW}$ to
  fourth order in $\Delta_{A,B,C,D}$  we obtain~\cite{charge,debanjan}
\begin{align}
\mathcal{F}_{\rm CDW}=&\alpha(|\Delta_A|^2+|\Delta_B|^2+|\Delta_C|^2+|\Delta_D|^2) \nonumber\\
&+\beta_0(|\Delta_A|^4+|\Delta_B|^4+|\Delta_C|^4+|\Delta_D|^4) \nonumber\\
&+\beta_1(|\Delta_A|^2+\Delta_B^2)(|\Delta_C|^2+|\Delta_D|^2) \nonumber\\
&+\beta_2 (\Delta_A\Delta_C\Delta_B^*\Delta_D^*+h.c.),
\label{o4}
\end{align}
where $\alpha = {\bar \alpha} (T - T_{\rm CDW})$,
 and
 ${\bar \alpha}={\bar \alpha} (T) \sim \Lambda/(v_xv_yT)$.

 The coefficients $\beta_i$ are obtained by evaluating square diagrams made out of fermions.
 It is straightforward to verify that $\beta_2$ is positive at all $T$. For such $\beta_2$,  the system
 favors the order with
a negative
$\Delta_A\Delta_C\Delta_B^*\Delta_D^*=-|\Delta_A\Delta_C\Delta_B^*\Delta_D^*|$.

We further notice that the Free energy is symmetric under $A\leftrightarrow B$, and $C\leftrightarrow D$ and that there are no additional couplings between $\Delta_A$ and $\Delta_B$ and between $\Delta_C$ and $\Delta_D$.  Accordingly, we set $|\Delta_A|=|\Delta_B|=|\Delta_y|$ and $|\Delta_C|=|\Delta_D|=|\Delta_x|$. The Free energy (\ref{o4}) then becomes
\begin{align}
\mathcal{F}_{\rm CDW}=&2\alpha(|\Delta_x|^2+|\Delta_y^2|)+2\beta_0(|\Delta_x|^4+|\Delta_y|^4) \nonumber\\
&+(4\beta_1-2\beta_2)|\Delta_x|^2|\Delta_y|^2+O(\Delta^6).
\end{align}
An elementary analysis then shows that
 CDW order is uni-directional when $2(\beta_1-\beta_0)>\beta_2$, and  bi-directional when $2(\beta_1-\beta_0) <\beta_2$.

The coefficients $\beta_i$ have to be computed along $\alpha =0$ line, i.e. for $T = T_{cdw} (x)$.  In practice, it is more convenient to keep $T$
 initially as a parameter and set $T = T_{cdw} (x)$ at a later stage.  At the lowest $T  \ll v_x \Lambda$, the
  coefficients $\beta_i$ have been  obtained previously~\cite{charge,debanjan}.    In this limit
\begin{align}
\beta_0&=\frac{1}{16\pi^2 v_x^2 v_y \Lambda},\nonumber\\
\beta_1&=\frac{\mu^2}{4\pi^2 v_x^2v_y\Lambda}\log\frac{v_x\Lambda}{T},~\beta_2=\frac{\mu^2}{16 v_x v_y T}.
\label{beta}
\end{align}
Clearly, at the lowest temperature,
$\beta_2\gg 2(\beta_{1}-\beta_0)$, i.e., the CDW is bi-directional.

 We extended the analysis of $\beta_i$ to higher temperatures.  Because $v_y \gg v_x$, there are two characteristic energy/temperature scales,
  $T_1 = v_x \Lambda$ and $T_2 = v_y \Lambda \gg T_1$.  Eq.\ (\ref{beta}) is valid for $T \ll T_1$.  At
   $T_2 \gg T \gg T_1$ we obtained, up to small corrections,
\begin{align}
\beta_0 &= C  \frac{\Lambda}{2T^2},~ \beta_2 = 2\beta_1 = C \mu^2 \frac{\Lambda}{T^2}
\label{beta_1}
\end{align}
where $C=7\zeta(3)/(16\pi^4 v_y)$ and
$\zeta(3)$ is the Riemann Zeta function.  We see that, again, $\beta_2- 2(\beta_{1}-\beta_0)$ is positive, i.e., the CDW is bi-directional.

At even higher temperatures $T \gg T_2$, we have
\begin{align}
\beta_0 &=(1 + \mu^4) {\tilde C} ,~ \beta_1 =\beta_2 = 2 \mu^2 {\tilde C} ,
\label{beta_2}
\end{align}
where $\tilde C=\Lambda^2/(192\pi^2 T^3)$.
In this situation
$\beta_2 - 2(\beta_1-\beta_0) \propto
(1 + \mu^4) - \mu^2 =
(1-\mu^2)^2 + \mu^2$.
 This is again positive, i.e., CDW order at the onset is again bi-directional.

We see therefore that CDW order at the onset is bi-directional for all $T$, when $T$ is considered as a parameter.  Obviously then, the CDW order is bi-directional along the whole $T_{\rm CDW} (x)$ line.

Although the structure of CDW order remains the same along $T_{\rm CDW}(x)$, the type of the order changes.  At the lowest temperature, $\beta_2$ is much larger than $\beta_1$ and $\beta_0$, and the CDW transition is first order.  In this situation, the analysis based on the comparison of coefficients of the quartic terms is, strictly speaking, incomplete, as one has to include higher order terms in $\Delta$ and analyze the structure of CDW order for finite $\Delta_x$ and $\Delta_y$ immediately below first-order transition.  At higher $T > T_1$, the CDW transition is second order and the analysis based on the comparison of the quartic terms is perfectly valid near the onset.

Before we proceed to include higher orders in $\Delta$'s, we note that the condition
\begin{align}
 \Delta_A\Delta_C\Delta_B^*\Delta_D^*=-|\Delta_A\Delta_C\Delta_B^*\Delta_D^*|
 \label{varphi}
 \end{align}
  actually holds
   for arbitrary magnitude of $\Delta$.
 To see this, we recall that the Free energy in Eq. (\ref{o4}) is obtained from the original model with fermion-fermion interaction by introducing $\Delta_{A,B,C,D}$ as the Hubbard-Stratonoivich fields, performing
  Hubbard-Stratononich transformation, integrating over fermions, and expanding in powers of $\Delta$.
    One can vary the relative phases between the $\Delta_{A,B,C,D}$ and minimize the Free energy before expanding in $\Delta$.
     The part of the Free energy that depends on the relative phases for hot spots $1,2,-1,-2$ is
\begin{align}
\mathcal {F}_\varphi=-\log\[\det \(
\begin{array}{cccc}
G_1^{-1}&\Delta_A&\Delta_C^*&0 \\
\Delta_A^*&G_2^{-1}&0&\Delta_D^*\\
\Delta_C& 0 & G_{-2}^{-1} & \Delta_B\\
0 & \Delta_D & \Delta_B^* &G_{-1}^{-1}
\end{array}\)
\],
\label{fphi}
\end{align}
where  $G_i=G(\omega_m,\epsilon_{i,k})=1/(i\omega_m-\epsilon_{i,k})$ is the Green's function.
 In Eq.\ (\ref{fphi}) the summation over $\omega_m$ and $k$ is assumed.

  In the bi-directional state we
 define
 $|\Delta_{A,B,C,D}|=|\Delta|$ and $\Delta_A\Delta_C\Delta_B^*\Delta_D^*=|\Delta|^4e^{i\varphi}$. We then expand the determinant in Eq.\ (\ref{fphi}) and obtain
 \begin{align}
 \mathcal {F}_\varphi= &-T\sum_{\omega_m,k}\log\[(\omega_m^2+\epsilon_1^2)(\omega_m^2+\epsilon_2^2) \right. \nonumber\\
& ~~~~~~~~~~~~\left.+4\omega_m^2|\Delta|^2 +2(1-\cos\varphi)|\Delta|^4\],
 \end{align}
 where we have used the fact that
 to linear order in momentum, counted from a hot spot,
 $\epsilon_{-i,k}=-\epsilon_{i,k}$.
 Minimizing $\mathcal {F}_\varphi$
  we obtain that $\varphi=\pi$,
  i.e.,  that
  $\Delta_A\Delta_C\Delta_B^*\Delta_D^*=-|\Delta_A\Delta_C\Delta_B^*\Delta_D^*|$.
The Free energy for $\Delta$ between
hot spots $3,4,-3,-4$ is analyzed in a similar way and the condition on the phase is the same $\varphi =\pi$.
 Hence
  the condition (\ref{varphi}) indeed minimizes the Free energy for arbitrary magnitudes of
  $\Delta_{A,B,C,D}$.
    This in turn allows us
    to fix the phase  before expanding in powers of $\Delta$.

\subsection{Uni-directional vs bi-directional order inside the CDW phase}~

\begin{figure}
\includegraphics[width=\columnwidth]{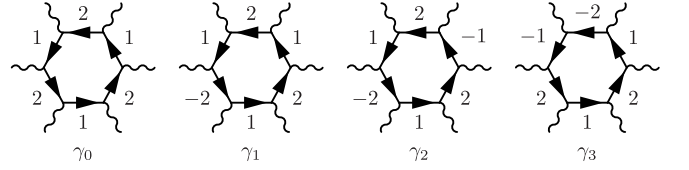}
\caption{Four types of six-leg diagrams, corresponding to $\gamma_{0,1,2,3}$. For simplicity we have only shown one diagram of each type.}
\label{fig_diag}
\end{figure}

We now analyze how the order changes inside the CDW phase.  For this, we extend the Free energy to include the terms of the sixth order in $\Delta$.
The full Free energy to this order is
\begin{align}
\mathcal{F}_{\rm CDW}=&\alpha(|\Delta_A|^2+|\Delta_B|^2+|\Delta_C|^2+|\Delta_D|^2) \nonumber\\
&+\beta_0(|\Delta_A|^4+|\Delta_B|^4+|\Delta_C|^4+|\Delta_D|^4) \nonumber\\
&+\beta_1(|\Delta_A|^2+\Delta_B^2)(|\Delta_C|^2+|\Delta_D|^2) \nonumber\\
& - 2 \beta_2 |\Delta_A\Delta_C\Delta_B^*\Delta_D^*|\nonumber\\
&+\gamma_0(|\Delta_A|^6+|\Delta_B|^6+|\Delta_C|^6+|\Delta_D|^6)\nonumber\\
&+\gamma_1[(|\Delta_A|^4+|\Delta_B|^4)(|\Delta_C|^2+|\Delta_D|^2)\nonumber\\
&~~~~~+(|\Delta_C|^4+|\Delta_D|^4)(|\Delta_A|^2+|\Delta_B|^2)]\nonumber\\
&+\gamma_2[|\Delta_A|^2|\Delta_B|^2(|\Delta_C|^2+|\Delta_D|^2)\nonumber\\
&~~~~~+|\Delta_C|^2|\Delta_D|^2(|\Delta_A|^2+|\Delta_B|^2)]\nonumber\\
&- 2\gamma_3(|\Delta_A|^2+|\Delta_B|^2+|\Delta_C|^2+|\Delta_D|^2)\nonumber\\
&~~~~~\times|\Delta_A\Delta_C\Delta_B^*\Delta_D^*| +O(\Delta^8),
\label{o6}
\end{align}
 and we already applied the condition on the relative phases, Eq. (\ref{varphi}).
 The prefactors for different $\Delta^6$ terms are obtained by evaluating six-leg fermion loop diagrams, which we show in Fig.\ \ref{fig_diag}.
  To simplify the calculations we  set $\mu=1$, i.e., assume a purely $d$-wave form factor for CDW. This simplifies the evaluation of the integrals but does not qualitatively affect the outcome.
  For $\mu =1$, the expressions for $\gamma_i$ are
 \begin{align}
 \gamma_0=&\frac{1}{3}\int G_1^3G_2^3+\frac{1}{3}\int G_3^3G_4^3\nonumber\\
 \gamma_1=&\int G_1^3G_2^2G_{-2}+\int G_3^3G_4^2G_{-4}\nonumber\\
  \gamma_2=&\gamma_3= \int G_1^2G_2^2G_{-1}G_{-2}+ \int G_3^2G_4^2G_{-3}G_{-4},
 \end{align}
 where
  the integration over $k$ and summation over $\omega_m$ are assumed.
The evaluation of these coefficients is standard but
the formulas are
  quite cumbersome and we refrain from presenting them.  As our primary goal is to understand what happens with
  bi-directional order as the magnitude of CDW order parameter gets larger,  we restrict with
  $T \ll v_x \Lambda$.
   In this limit  one can safely extend the upper limit of momentum integration to
    infinity.
    Both $\gamma_0$ and $\gamma_1$ then vanish  due to triple poles in the integrands.
On the other hand, the momentum integral for  $ \gamma_{2}=\gamma_3$ contains poles in different momentum half-planes
and hence remains finite. This integral diverges in the infrared, and the divergence is cut by $T$.
 An explicit calculation shows that
 $\gamma_2=\gamma_3$ is negative:
\begin{align}
\gamma_2=\gamma_3=-\frac{1}{768v_xv_yT^3}.
\label{gamma}
\end{align}

The Free energy is again symmetric under $A\leftrightarrow B$, and $C\leftrightarrow D$ and we set $|\Delta_A|=|\Delta_B|=|\Delta_y|$ and $|\Delta_C|=|\Delta_D|=|\Delta_x|$.
Neglecting $\beta_1$ compared to $\beta_2$ (recall
that $\beta_1 \ll \beta_2$ at $T \ll v \Lambda$, see Eq. \ref{beta})
 and   keeping only $\gamma_2$ and $\gamma_3$ terms in (\ref{o6})
  we simplify the Free energy to
\begin{align}
\mathcal{F}_{\rm CDW}=&2\alpha(|\Delta_x|^2+|\Delta_y|^2) + \beta_0 (|\Delta_x|^4+|\Delta_y|^4) \nonumber\\
-&2\beta_2|\Delta_x|^2|\Delta_y|^2 +4|\gamma_2||\Delta_x|^2|\Delta_y|^2(|\Delta_x|^2+|\Delta_y|^2).
\label{ch_1}
\end{align}

Because
 $\beta_2 \gg \beta_0$ at $T\ll v\Lambda$,  the system initially develops a bi-directional CDW order via a first-order transition
at a positive
 $\alpha_{cr} = {\beta_2^2}/({32\gamma_2})$.   At the same time, the sign of the sixth-order term is opposite to that of the quartic term, hence, the energy gain associated with bi-directional order gets weaker as
 the magnitude of $\Delta_{x,y}$ grows.   When the sixth-order term gets larger than the
  quartic term,
  it becomes energetically advantageous for the system to
   switch to a uni-directional order.
  Comparing the Free energies (\ref{ch_1}) of uni-directional and bi-directional CDW orders
   we find
   that the transition from bi-directional to uni-directional order occurs  approximately at
$\alpha = 0$.
We have explicitly verified that, within the range $0<\alpha<
\alpha_{cr}$,
 the quartic term $-2\beta_2|\Delta_x|^2|\Delta_y|^2$, which favors bi-directional order is
 larger by magnitude
 than the sixth order term $4|\gamma_2||\Delta_x|^2|\Delta_y|^2(|\Delta_x|^2+|\Delta_y|^2)$, which favors uni-directional order.
This implies that, despite that the CDW transition is first-order,
 the system still initially develops bi-directional order and only later (at larger $\Delta$) the order switches to uni-directional.

\subsection{The CDW phase diagram}
 We use the results from the previous section to construct the phase diagram.
  The mean-field phase diagram is shown 
  in the inset of Fig.\ 1.  Upon lowering temperature or doping,
 the system first develops a bi-directional CDW order
  via
   a second-order transition at a higher $T_{\rm CDW} $ and via first-order transition at a lower $T_{\rm CDW}$.
     However,
     CDW order  goes back to the uni-directional as the system moves some distance into the CDW-ordered phase.
     As the consequence,  uni-directional order exists in the large potion of the
      CDW phase.
       We  did not extend CDW region in Fig.\ 1 down to $T=0$.  At $T \to 0$, the prefactors of $\Delta^4$, $\Delta^6$, etc terms diverge with progressively higher powers of $1/T$ and one has to analyze the interplay between uni-directional and bi-directional orders without performing the Landau expansion.

The transformation from uni-directional and bi-directional CDW order at higher dopings has been detected in Refs.\ \onlinecite{keimer,andrea}.  Ref.\ \onlinecite{keimer} found that the order goes back to uni-directional inside the CDW phase. Our results are fully consistent with these data.

 Beyond mean-field, the interplay between uni-directional and bi-directional CDW is influenced by the additional nematic transition which occurs above the temperature at which uni-directional CDW order sets in and breaks $C_4$ rotational symmetry down to $C_2$.
  The microscopic rationale for the existence of such transition has been presented before, both for the cuprates~\cite{charge} and Fe-pnictides~\cite{rafael_nem}, and we do not repeat it here.
  The  feedback effect from the nematic transition on the  primary CDW order increases the onset temperature of the uni-directional order compared to that in the mean-field
   approximation. Once the nematic transition temperature $T_{\rm nem}$ exceeds $T_{\rm CDW}$, bi-directional order either gets completely eliminated or the magnitudes
     of $\Delta_x$ and $\Delta_y$ within bi-directional phase become non-equal as the consequence of the broken $C_4$ symmetry. To distinguish between the two possibilities,  one needs to compute Free energies beyond mean-field, which is beyond the scope of the current paper. On general grounds,  the condition
       $T_{\rm nem} > T_{\rm CDW}$ is unlikely to be satisfied at small $T_{\rm CDW}$ because there the difference between the Free energies of the
       uni- and bi-directional phases immediately below $T_{\rm CDW}$
 are the largest, but it well may get satisfied at higher $T_{\rm CDW}$, when the Free energy difference between the two phases
 right below $T_{\rm CDW}$
 gets smaller.  We show the phase diagram beyond mean-field in Fig.\ 1, assuming that bi-directional phase gets eliminated once $T_{\rm nem}$ exceeds $T_{\rm CDW}$.

Another dashed line in Fig.\ 1  marks the temperature at which the system breaks time-reversal symmetry.
  This line lies on top of both uni-directional and bi-directional phases.  For the uni-directional phase, its presence is associated with the fact that  CDW orders with the same ${\bf Q}$ but opposite center-of-mass momentum, e.g.,  $\Delta_{A}$ and $\Delta_{B}$, are un-coupled within the hot-spot model but
   become linearly coupled via a term $\sim\Delta_A\Delta_B^*+h.c$ in a more generic model in which CDW coupling is extended to  fermions away from hot spots.
    A model calculation (see Ref.\ \onlinecite{charge,pdw}) have found that the relative phases between $\Delta_{A}$ and $\Delta_{B}$ are locked at $\pm \pi/2$.
      The selection of $\pi/2$ or $- \pi/2$  breaks the $\z$ symmetry.  Because $\Delta_A$ and $\Delta_B$ transform into each other under time-reversal,
       the selection $\pi/2$ or $- \pi/2$ implies the breaking of  time-reversal symmetry.
     In the bi-directional phase, the phases of the order parameters $\Delta_A$, $\Delta_B$, $\Delta_C$, and $\Delta_D$ are locked by $ \Delta_A\Delta_B^*\Delta_C\Delta_D^*=-|\Delta_A\Delta_B^*\Delta_C\Delta_D^*|$ [Eq.\ (\ref{varphi})].
     Once the system selects the relative phase between $\Delta_A$ and $\Delta_B$  to be $\pi/2$ or $-\pi/2$, the phase difference between $\Delta_C$ and $\Delta_D$ is adjusted
     to be the same as between  between $\Delta_A$ and $\Delta_B$.  This implies that the time-reversal symmetry breaking does not differentiate between uni-directional and bi-directional CDW orders.

     The existence of the nematic phase above the left half of the CDW dome has been confirmed by in-plane resistivity measurements~\cite{taillefer_new}.
       It would be interesting to compare its location with the onset of the Kerr effect~\cite{kerr} and the intra-unit-cell order observed in neutron scattering~\cite{sidis}, which both detect time-reversal symmetry breaking.  From the theoretical perspective, the critical temperature of time-reversal symmetry breaking can be either higher or lower than $T_{\rm nem}$, depending on model parametetrs.

\section{The effect of the PDW order}
In this section we discuss the  structure of CDW order in a situation when CDW order develops along with PDW order, or when PDW order, of one kind or another,
 develops before CDW order.  The PDW order is a superconducting order with a non-zero total momentum of a pair (like in Fulde-Ferrel-Larkin-Ovchinnikov (FFLO) state, but in zero field).
Signatures of the PDW order has  been  detected in the tunneling experiments on the cuprates~\cite{davis_last}, and PDW order has been obtained in various analytical~\cite{patrick,agterberg,kivelson,agterberg2,berg,pdw_frad,pdw,pdw2,pepin,chan} and numerical~\cite{troyer} theoretical calculations.  It was argued that
the presence of a PDW order
explains several experimental features in the pseudogap phase~\cite{patrick,pdw2,agterberg,kivelson}.

\subsection{PDW order within the spin-fermion model}
Within the
 spin-fluctuation scenario, one can introduce two different kinds of PDW order. One
 connects the pairs of hot spots that are separated in momentum by $Q_x$ or $Q_y$, like CDW order does.  The corresponding  PDW order parameters are, e.g., $\bar\Delta_A\sim i\sigma^y_{\alpha\beta}c_{1\alpha}(\tilde k)c_{2\beta}(-\tilde k)$, where $\tilde k$ is the momentum deviation from the corresponding hot spot.
 The PDW order of this kind is a ``partner" of CDW in the same way as CDW order with diagonal momenta $(Q,\pm Q)$ is a partner of magnetically-mediated $d$-wave superconductivity~\cite{ms,efetov}. The partnership means that the two orders (PDW and CDW with $Q_x$ and $Q_y$ in our case) are degenerate in the ``hot spot only" model due to underlying  SU(2) particle-hole symmetry~\cite{ms}.
 The symmetry between CDW and PDW orders enlarges the order parameter manifold for each hot spot pair from U(1) to SO(4).

An SO(4)-covariant Ginzburg-Landau Free energy that incorporates both CDW and PDW components has been derived and analyzed  in Refs.\ \onlinecite{pdw,pdw2}.
  For completeness, we briefly review the results here. 
 
 In the presence of PDW, the  U(1) CDW order parameter, say $\Delta_A$, is replaced by a $2\times2$ matrix which has both CDW and PDW components
\begin{align}
{\bf\Delta}_A\equiv\(\begin{array}{cc}
\bar\Delta_A &\Delta_A^*\\
-\Delta_A & \bar\Delta_A^*
\end{array}\).
\end{align}
This $\so$ order parameter ${\bf \Delta}_A$ couples to particle-hole doublets $\Psi_1(k)=(c_{1\uparrow}(k),c^\dagger_{1\downarrow}(-k))^T$ and $\Psi_2(k)=(c^\dagger_{2\downarrow}(-k),c_{2\uparrow}(k))^T$ via
\begin{align}
\mathcal{H}'_{{\bf \Delta}_A}=\Psi^\dagger_{1\mu}{\bf\Delta}_A^{\mu\nu}\Psi_{2\nu}.
\end{align}
The other CDW/PDW order parameters ${\bf\Delta}_{B,C,D}$ can be similarly defined and coupled to fermions.
 The 
 Free energy 
 is quite similar to that in Eq.\ (\ref{o4}):
 \begin{align}
 &\mathcal{F}_{\rm CDW/PDW}\nonumber\\
 =&\alpha \Tr({\bf \Delta}_A^\dagger {\bf \Delta}_A+{\bf \Delta}_B^\dagger {\bf \Delta}_B+{\bf \Delta}_C^\dagger {\bf \Delta}_C+{\bf \Delta}_D^\dagger {\bf \Delta}_D)\nonumber\\
&+\beta_0\Tr({\bf \Delta}_A{\bf \Delta}_A^\dagger{\bf \Delta}_A{\bf \Delta}_A^\dagger+{\bf \Delta}_B{\bf \Delta}_B^\dagger{\bf \Delta}_B{\bf \Delta}_B^\dagger\nonumber\\
&+{\bf \Delta}_C{\bf \Delta}_C^\dagger{\bf \Delta}_C{\bf \Delta}_C^\dagger+{\bf \Delta}_D{\bf \Delta}_D^\dagger{\bf \Delta}_D{\bf \Delta}_D^\dagger)\nonumber\\
&+\beta_1\Tr\[({\bf \Delta}_A{\bf \Delta}_A^\dagger+{\bf \Delta}_B{\bf \Delta}_B^\dagger)({\bf \Delta}_C{\bf \Delta}_C^\dagger+{\bf \Delta}_D{\bf \Delta}_D^\dagger)\]\nonumber\\
&+\beta_2\[\Tr({\bf \Delta}_A^\dagger{\bf \Delta}_B{\bf \Delta}_C^\dagger{\bf \Delta}_D)+h.c.\]. \nonumber\\
 \end{align}

 The structure of the full CDW/PDW order is also quite similar to that for a pure CDW order. Namely,
the CDW/PDW order
  can be either uni-directional (state I), when CDW/PDW develops either on bonds $A,C$ or on bonds $B,D$ (see Fig.\ 1), or bi-directional (state II),
  when CDW/PDW  develops for all bonds $A,B,C,D$. The CDW/PDW states I and II are  the counterparts of the uni-directional and bi-directional pure CDW order,
    and the selection of state I or state II is determined by the same interplay between the coefficients $\beta_0$, $\beta_1$ and $\beta_2$ as in the previous section. Borrowing the results, we argue that CDW/PDW order is bi-directional at the onset.  The structure of the CDW/PDW order inside the ordered phase is again determined by the interplay between terms of order $\Delta^4$ and of order $\Delta^6$.  We extended the analysis of the $\Delta^6$ terms in the previous section to  SO(4) CDW/PDW model and found the same result as earlier, namely that the order changes to uni-directional inside the CDW/PDW state. This implies that the phase diagram of the  SO(4) CDW/PDW model is
     at least qualitatively
     the same as for the pure CDW order, see Fig.\ 1.

A more subtle issue is the distribution of CDW and PDW order parameters in the CDW/PDW state, particularly in the bi-directional state, where the combined CDW/PDW order develops with both $Q_x$ and $Q_y$.  The  Landau functional for SO(4) CDW/PDW  order parameter is highly degenerate as for each bond the system can  develop an arbitrary ``mixture" of CDW and PDW. The degeneracy gets broken when one includes into consideration the fact that
CDW order with, say, $Q_x$ and PDW order with orthogonal $Q_y$ generate a secondary homogenous superconducting order, and this gives rise to additional lowering of the Free energy (see Ref.~\onlinecite{pdw2}). As a result,  the true ground state for State II is the one for which CDW only forms along one bond direction, say $(A,B)$, while PDW only forms along the other bond direction, $(C,D)$.  Such a state breaks $C_4$ lattice rotational symmetry down to $C_2$ and for CDW order, such a state is still uni-directional in the sense that CDW only develops with either $Q_x$ or $Q_y$.  Then, x-ray experiments, which only probe CDW order, should not detect any changes with doping, despite that the full CDW/PDW order becomes bi-directional.

Note in passing
 that in a more general analysis,
 which (i) includes the Fermi surface curvature into the dispersion and (ii) goes beyond the hot spot model,
  the CDW and PDW are not degenerate, but remain strong competitors~\cite{pdw}.
  The inclusion of the Fermi surface curvature favors
  PDW order whose mean-field onset temperature becomes higher than that for  CDW order.
    The extension of the model beyond hot spots, on the other and,
    favors CDW order due to feedback from
 the time-reversal symmetry breaking  at a higher $T$.  Such an order does not develop for PDW.
  The analysis in Refs.\ \onlinecite{pdw,pdw2} shows that, in the bi-directional CDW/PDW state,  the system either develops a pure bi-directional CDW or PDW order, or develops an order with CDW
  along one bond
   direction   and PDW along the
other direction.

\begin{figure}
\includegraphics[width=0.35\columnwidth]{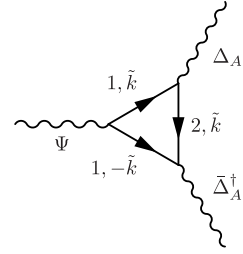}
\caption{The triangular diagram representing the coupling between CDW/PDW orders
 for the same
 bond
  connecting hot spots
  (bond $A$ in this case) and the Amperean pairing order $\Psi$, which involves
  two
  fermions with approximately the same momenta. $\tilde k$ is the momentum deviation from a hot spot.}
\label{amp}
\end{figure}

\subsection{PDW order from Amperean pairing}

 Another type of PDW order was originally introduced in Ref.~\onlinecite{patrick} in the framework of strong-coupling (Mott) scenario~\cite{mott} and was termed as ``Amperean pairing". This PDW order involves fermions with close momenta
 ${\bf k} \pm \delta
 {\bf k}$,
  such that the total momentum of a pair is
  $2{\bf k}$.
 It was later re-introduced for a  hot spot model~\cite{agterberg} and was shown, among other things,  to give rise to the breaking  of
 time-reversal symmetry.
 In Ref.\ \onlinecite{agterberg}, hot spots ${\bf k}_h$ were not identified precisely with the crossing points of fermionic dispersion and magnetic Brillouin zone boundary, and $2{\bf k}_h$ was set to be a generic momentum $(Q_1, Q_2)$.
 In our spin-fluctuation model, hot spots are of magnetic origin and
 the hot spot momenta are ${\bf k}_h = (k, \pi\pm k)$. Accordingly, the total momentum of a pair $2{\bf k}
 =2{\bf k}_h = (2k, \pm 2k)=(Q,\pm Q)$ is along one of the two Brillouin zone diagonals.
 We label such pair-density-wave order as
 PDW$^*$.

  An interesting situation develops when PDW$^*$, with diagonal momentum $(Q,Q)$, is present along with the bi-directional CDW/PDW order.
    In terms of hot spots, PDW$^*$ introduces a term $\sim i\sigma_{\alpha\beta}^y c_{i,\alpha}^\dagger(\tilde k) c_{i,\beta}^\dagger(-\tilde k)$ into the fermionic dispersion. In the presence of such a term,  the particle and  the hole at a given hot spot
      get mixed.
      As the consequence,  the CDW order in particle-hole channel and the PDW order in particle-particle channel get hybridized, and the development of
      of  one immediately generates the other,
 i.e., the PDW order along a given bond  induces  CDW order along the same bond and vice versa.
  Mathematically, this hybridization is reflected in the fact that CDW with $Q_x = (Q,0)$, PDW with $Q_y = (0,Q)$ and PDW$^*$ (which we denote as $\Psi$) with
     $(Q,Q)$ can be combined into a triangular diagram shown in Fig.\ \ref{amp}.
     This triple diagram generates the term in the Free energy which is bi-linear in CDW and PDW orders.  As a result, if we
      define CDW component along a particular bond as $\Delta \cos \theta$
    and PDW component along the same bond as $\Delta \sin \theta$, the Free energy becomes
    \begin{align}
\mathcal{F}_{\theta}=& A \sin {2\theta} + B \sin^2{2\theta} + ...
\end{align}
where $A$ is proportional to the magnitude of  PDW$^*$ order $\Psi$. Minimizing with respect to $\theta$ we immediately obtain in equilibrium $\sin 2\theta = -A/(2B)$, which implies that both CDW and PDW are present along each direction, but the magnitude of CDW order in one direction is not equivalent to that in the other direction.
 This is consistent with x-ray experiments, which at larger doping found non-equal $\Delta_x$ and $\Delta_y$ in every domain.

The hybridization between CDW and PDW orders in our case is quite similar to that between singlet and triplet pairing channels either in the context of spin-orbit coupling~\cite{gorkov} or in the spin-density-wave state of the Fe-pnictides~\cite{alberto}, when spin is no longer a conserved quantum number.

\section{Summary}

In this work we adopted the spin-fluctuation formalism and analyzed in detail  the interplay between uni-directional and bi-directional charge orders with axial momenta $Q_x$ and $Q_y$ in the cuprates.
We derived the Landau Free energy to sixth order in CDW order parameters $\Delta_x$ and $\Delta_y$.
 These two order parameters couple to fermions in hot regions on the FS, and the prefactors in the Landau Free energy are
   obtained by evaluating
    loop diagrams made out of hot fermions.
We found that the CDW order is  bi-directional at its onset, but changes to
  uni-directional inside the CDW-ordered phase, once the magnitude of the order parameter exceeds some critical value.  This is consistent with recent X-ray data on YBCO~\cite{keimer,andrea}.

  We also discussed the effect of a PDW order. An axial PDW order also emerges from the spin-fluctuation scenario and is degenerate with the axial CDW order in the hot spot model, due to particle-hole SU(2) symmetry.  Within this model, the bi-directional state  is actually uni-directional for CDW as it only develops with $Q_x$ or $Q_y$, the order along the orthogonal direction is PDW.  We analyze the case when, in addition to axial CDW/PDW, the system also develops, by different reasons, an Amperean PDW with diagonal momentum $(Q,Q)$. We found that Amperean PDW couples axial CDW and PDW along each bond. As a result, in the bi-directional state, CDW order  develops on each bond and hence by itself becomes bi-directional. Then uni-directional and bi-directional CDW/PDW states show different behavior already in the experiments like X-ray, which at present probe only CDW order.

  The issue which we did not address in this work is the relation to quantum oscillation experiments. These experiments were interpreted as evidence for CDW-induced electron pockets, and this interpretation implies that CDW order is bi-directional~\cite{suchitra,deban_qo}, even in the doping range where x-ray measurements report uni-directional order.  The apparent contradiction can be resolved if it turns out that a magnetic field, in which quantum oscillation measurements have been performed, pushes the system towards bi-directional order.  This, however, needs to be verified in explicit calculations.

Finally, we note that the transformation from bi-directional to uni-directional order (i.e., from checkerboard to stripe order) inside the ordered phase
is not specific to the cuprates and has recently been observed and analyzed in  iron-based superconducting materials~\cite{c4_pnic}.
  This is yet another evidence that the two families of materials have much in common.

\begin{acknowledgments}
We thank D. F. Agterberg, A. Damascelli, D. Chowdhury, R. Fernandes, E. Fradkin, B. Keimer, and especially Jian Kang for fruitful discussions. The work was supported by the NSF DMR-1523036 (YW and AC) and by the Gordon and Betty Moore Foundation's EPiQS Initiative through Grant No.\ GBMF4305 at the University of Illinois (YW).
\end{acknowledgments}

\end{document}